\documentstyle[sprocl]{article}
\input{psfig.sty}
\def\Journal#1#2#3#4{{#1} {\bf #2}, #3 (#4)}
\def\CMP{\em Comm. Math. Phys.}
\def\JMP{\em J. Math. Phys.}
\def\GRG{\em Gen. Rel. Grav.}

\def\be{\begin{equation}}
\def\ee{\end{equation}}
\def\bea{\begin{eqnarray}}
\def\eea{\end{eqnarray}}

\newtheorem{lemma}{Lemma}
\newtheorem{proposition}{Proposition}

\begin{document}

\title{IDEAL GAS STEPHANI UNIVERSES}
\author{B. COLL }
\address{Syst\`emes de ref\'er\'ence spatio-temporels, CNRS\\
DANOF, Observatoire de Paris\\
F-75014 Paris, France.\\
E-mail: bartolome.coll@obspm.fr}
\author{J.J. FERRANDO}
\address{Departament d'Astronomia i Astrof\'{\i}sica, Universitat
de Val\`encia,
\\ E-46100 Burjassot, Val\`encia, Spain.\\
E-mail: joan.ferrando@uv.es}
% You may repeat \author \address as often as necessary

\maketitle\abstracts{The Stephani Universes that can be interpreted as
an ideal gas evolving in local thermal equilibrium are determined, and
the method to obtain the associated thermodynamic schemes is given.}

\section{Introduction}

The conformally flat perfect fluid solutions of Einstein equations
with nonzero expansion were considered by Stephani \cite{st} and
are usually called Stephani Universes.
These solutions were rediscovered by Barnes \cite{ba} who
obtained them
as the conformally flat class of irrotational and shear-free perfect
fluid spacetimes with nonzero expansion.

In order to generalize the cosmological principle, Bona and Coll
\cite{bc} looked for spacetimes admitting an iso-invariant
synchronization,
and without any hypothesis on the energy tensor, they likewise found
the Stephani Universes.

An iso-invariant synchronization implies that an irrotational and
shear-free observer $u$ with non zero expansion exists, and the induced
metric on the orthogonal hypersurfaces are of constant curvature. Then,
there exists an adapted coordinate system, $u= \frac{1}{\alpha}
\partial_t$, such that the line element writes:
\be
\begin{array}{c}
ds^2 = -\alpha^2 dt^2 + \Omega^2 \delta_{ij} dx^i dx^j
\\
\displaystyle{\Omega \equiv \frac{R(t)}{1+ 2 \vec{b}(t) \cdot \vec{r} +
\frac{1}{4} K(t) r^2}}, \qquad  \qquad
\alpha \equiv R \partial_R \ln \Omega
\end{array}
\label{eq:metric}
\ee
$R(t)$, $\vec{b}(t)$ and $K(t)$  being 5 arbitrary functions of time.

The metric (\ref{eq:metric}) is a perfect fluid
solution, $T = (\rho+p)u \otimes u +pg$, with energy density and
pressure given by
\be
\rho = {3 \dot{R}^2 \over R^2} + {3 \over R^2} (K-4b^2) \; ,
\qquad  \qquad
p= - \rho - {3 \over R} {\partial_R\rho \over \alpha}
\label{eq:dp}
\ee
Moreover, the expansion of $u$ and the curvature of the spatial metric
are homogeneous, and they are given, respectively, by
\be
\theta(t) = {3 \dot{R} \over R} \not= 0 \; ,   \qquad    \qquad
\kappa(t)= {1 \over R^2} (K-4b^2)
\ee

Stephani Universes are perfect fluid solutions of Einstein equations,
but what is the physical meaning of these perfect fluids?
Bona and Coll \cite{bc} gave the first step to answer this question:
they studied when these perfect fluid solutions admit a thermodynamic
scheme, and they showed that the thermodynamic Stephani Universes admit
at least a three-dimensional group of isometries. This topic
has been again considered later \cite{ks} and, in a recent paper,
Sussman \cite{su} analyzes a family of spherically symmetric Stephani
Universes that may be interpreted as either a classical monoatomic ideal
gas or a matter-radiation mixture.

Here we determine {\it all} Stephani Universes that represent the
evolution in local thermal equilibrium of a (generic) ideal gas. In
section 4 we present the main results that are based on previuos ones:
firstly, on the above cited
work by Bona and Coll \cite{bc} about the thermodynamic
schemes in this class of solutions, that we shorten here in
section 2; secondly, on our hydrodynamic characterization of an ideal
gas
in local thermal equilibrium \cite{cf}, that we summarize in section 3.

\section{Thermodynamic Stephani Universes}

The usual physical interpretation of a perfect fluid follows
when it has a conservative evolution in local thermal equilibrium
(l.t.e.). This means:
\begin{itemize}
\item
Energy-momentum conservation:
$\ \ \nabla \cdot T = 0$.
\item
The energy density $\rho$ is decomposed in terms of the matter density
$r$ and the specific internal energy $\epsilon$:
$\ \ \rho= r(1+\epsilon)$.
\item
The equation of conservation of matter is required:
$\ \ \nabla \cdot (ru) = 0 $.
\item
The thermodynamic variables are related by equations of state compatibles with the thermodynamic relation:
$\ \ T ds =  d \epsilon + p d (1/r) $.
\end{itemize}
Bona and Coll \cite{bc} looked for the Stephani Universes admitting the
above
thermodinamic scheme. The results that will be usefull for us here can
be summarized in the following two lemmas. \\
\begin{lemma}
A Stephani Universe admits a barotropic thermodynamic scheme iff it admits a six-dimensional isometry group.
\end{lemma}
\vspace{0.2cm}
So, this case corresponds to the Friedmann-Robertson-Walker limit that occurs when $K=constant$ and $\vec{b}=constant$.
\\
\begin{lemma}
A Stephani Universe admits a strict (non barotropic) thermodynamic scheme iff it admits a three-dimensional isometry group.

For the thermodynamic Stephani Universes, the metric may be written
$$
ds^2 = -\alpha^2 dt^2 + \Omega^2 \delta_{ij} dx^i dx^j
$$
\be
\Omega \equiv \frac{w}{z} L, \qquad  \qquad
\alpha \equiv R \partial_R \ln L                  \label{eq:termetric}
\ee
$$L \equiv \frac{R(t)}{1+ K(t) w}, \qquad  \qquad
w \equiv \frac{z}{1 + {\epsilon \over 4}r^2}$$
$R(t)$ and $K(t)$ being two arbitrary functions of time.
\end{lemma}
\vspace{0.2cm}
The symmetry group is spherical, plane or pseudospherical depending on
$\epsilon$ to be $1$, $0$ or $-1$.

For the thermodynamic perfect fluid solutions (\ref{eq:termetric}) the
expressions (\ref{eq:dp}) for the energy density and pressure
become
\be
\rho = {3 \dot{R}^2 \over R^2} + {3 \over R^2} (\epsilon - K^2) \; ,
\qquad  \qquad
p= - \rho - {3 \over R} {\partial_R\rho \over \alpha}  \label{tdp}
\ee

\section{Ideal gas evolving in local thermal equilibrium}

The definition of local thermal equilibrium given at the begining of the
previous section introduces thermodynamic
variables, like $r$, $\epsilon$, $s$, and $T$, that are not present in
the perfect fluid energy tensor. Then, a natural question arises: does
an equivalent formulation of the l.t.e. exist that
reduces to the addition to the energy-momentum conservation equation
of a new condition involving only the hydrodynamic variables $(u,
\rho,p)$?. We gave years ago \cite{cf2} a positive answer to this
question:\\
\begin{lemma}
A divergence-free perfect fluid
energy tensor evolves in l.t.e. if, and only if, the space-time
function $\chi \equiv \dot{p}/\dot{\rho}$,
called the indicatrix of l.t.e.,
depends only on the variables $p$ and $\rho$ :
$d\chi \wedge dp \wedge d\rho = 0$
\end{lemma}
\vspace{0.2cm}

This result has the conceptual interest of giving an exclusively
hydrodynamic version of l.t.e.. But moreover it also
has a practical utility because provides a deductive algorithm to
detect wether or not a given divergence-free perfect fluid energy tensor
evolves in l.t.e.. In the Spanish Relativity Meeting-96
\cite{cf3} we
pointed out these applications and gave a concrete exemple studying the
thermodynamic class II Szekeres-Szafron space-times.

Nevertheless, in order to have a more accurate physical meaning of a
perfect fluid it will be usefull to study the indicatrix function of
several particular media. Elsewhere \cite{cf} we have given this
hydrodynamic
characterization for some representative cases.
We present below in two propositions
a part of our results about ideal gases that we will use in next
section.

A (generic) ideal gas satisfies the equation of state $p=krT$. For it
we have the following hydrodynamic characterization \cite{cf}:
\vspace{0.1cm}
\begin{proposition}
The necessary and sufficient condition for a non barotropic and non
isoenergetic ($\dot{\rho} \not =0$) divergence-free perfect
fluid energy tensor $(u,\rho,p)$ to represent the l.t.e. evolution of an
ideal gas is that the indicatrix function $\chi\equiv
\dot{\rho}/\dot{p}$ be a function of the variable $\pi \equiv p/\rho$,
$\chi=\chi(\pi) \not= \pi$:
\begin{equation}
d \chi \wedge d\pi = 0 , \qquad \chi \not= \pi   \label{xi-pi}
\end{equation}
\end{proposition}

\begin{proposition}
A non barotropic and non isoenergetic divergence-free perfect
fluid energy tensor $(u,\rho,p)$ verifying (\ref{xi-pi}) represents
the l.t.e. evolution
of the ideal gas with specific internal energy
$\epsilon$, temperature $T$, matter density $r$, and specific entropy
$s$ given by
\bea
\epsilon(\rho,p) = \epsilon(\pi) \equiv e(\pi)-1;
\qquad      \qquad
T(\rho,p) = T(\pi) \equiv {\pi \over k} e(\pi) \label{T} \\
r(\rho,p) = {\rho \over e(\pi)}; \qquad      \qquad
s(\rho,p) = s_0 + k [\int \phi(\pi) d\pi - {\rm ln}\rho]   \label{r-s}
\eea
$e(\pi)$ and $\phi(\pi)$ being, respectively,
\begin{equation}
e(\pi) = e_0 e^{\int_0^{\pi}\psi(\pi)d\pi}, \quad
\psi (\pi) \equiv \frac{\pi}{(\chi(\pi)-\pi)(\pi+1)};
\qquad  \phi(\pi) \equiv {1 \over \chi(\pi)-\pi}
\label{e-pi}
\end{equation}
\end{proposition}
\vspace{0.1cm}
The above two propositions provide an algorithm to detect, in any
given family of divergence-free perfect fluids ${\bf T} =
\{T \equiv [u^{\alpha}(x^{\beta}), \rho(x^{\beta}), p(x^{\beta})]
\}$,
those that represent an ideal gas evolving in local thermal equilibrium:
\begin{enumerate}
\item
Calculate the coordinate dependence of the space-time functions
$\ \dot{p}/\dot{\rho} = \chi(x^{\beta})$ and $\  p/\rho
=\pi(x^{\beta})$
\item
Determine the gas ideal subset of ${\bf T}$ by imposing the
hydrodynamic condition of the proposition 1: $\ \  d \chi \wedge d \pi
= 0$.
\item
Obtain the explicit expression of the indicatrix function:
$\ \ \chi=\chi(\pi)$.
\item
Calculate, from $\chi(\pi)$, the functions $e(\pi)$ and
$\psi(\pi)$ given in (\ref{e-pi}), and obtain the thermodynamic
variables using (\ref{T}) and (\ref{r-s}).
\end{enumerate}

\section{Ideal gas Stephani solutions}

In order to determine the ideal gas Stephani Universes we start from the
thermodynamic ones (\ref{eq:termetric}), and we must study the
restrictions that our ideal gas characterization imposes on the
functions $R(t)$ and $K(t)$. We follow the algorithm presented at the
end of previous section:

\subsection*{1. Variables $\chi$ and $\pi$}

From (\ref{tdp}) and
(\ref{eq:termetric}), a direct calculation lends to:
\bea
p/\rho =\pi(R, w)= \frac{a(1+Kw)}{1+(K-RK')w}-1, \qquad
a=a(R)\equiv - \frac{R \rho'}{3 \rho}  \ \ \  \label{eq:pi}  \\
\dot{p}/\dot{\rho} = \chi(\pi,R) =  \pi + \frac{1}{3} -
\frac{1}{3}(\pi+1)[\frac{a'R}{a^2}+\frac{1}{a}+(\pi+1-a)\frac{RK''}{a^2K'}]
\ \ \label{eq:chi}
\eea
where the prime indicates derivative with respect to the variable $R$.

\subsection*{2. Ideal gas hydrodynamic condition: $\; d \chi \wedge d
\pi = 0$}

This condition restricts, in a first step, the functions $\rho(R)$ and
$K(R)$ that turn out to be submitted to the second order differential
equations
\bea
a'' - \frac{2a'^2}{a} = a' \frac{K''}{K'}   \label{e1} \\
K'''-K''(\frac{K''}{K'}-\frac{1}{R})=2K''\frac{a'}{a}  \label{e2}
\eea
Secondly, we put every solution $\rho(R)$, $K(R)$ for these equations in
the expression (\ref{tdp}) of the energy density $\rho$ and we obtain
a generalized Friedmann equation for the expansion factor $R(t)$:
\be
\rho(R) = {3 \dot{R}^2 \over R^2} + {3 \over R^2} [\epsilon - K^2(R)]
\ee

\subsection*{3. Indicatrix function: $\chi=\chi(\pi)$}

At this point we look for the functional expression to the indicatrix
function. Assuming equations (\ref{e1}) and (\ref{e2}), the indicatrix
(\ref{eq:chi}) becomes
\be
\chi(\pi) = \alpha \pi^2 + \gamma \pi + \beta, \qquad \qquad \alpha +
\beta = \gamma - 2/3    \label{chipi}
\ee

\subsection*{4. Thermodynamic variables}

Finally, the form (\ref{chipi}) of the indicatrix function determines,
using the results in proposition 2, the other thermodynamic variables,
and so, the properties of the associated thermodynamic scheme.
%\section{Some comments}
\vspace{0.1cm}\\

In contrast to the Sussman \cite{su} intuitive approach leading to very
partial results, our algorithm gives directly, and in the
above simple
form, the {\em whole} class of ideal gas Stephani Universes.
After that, we may look for a more specific ideal gases by imposing the
corresponding restrictions to the indicatrix function (\ref{chipi}).
For example, if
one collapses the whole space of solutions of equations
(\ref{e1}) (\ref{e2}) to the particular
solutions $a(R) = Constant$, and one imposes the gas to be monoatomic,
one obtains the partial results by Sussmann.
The physical meaning of the other cases will be considered elsewhere.

\section*{Acknowledgments}
This work has been supported by the Spanish DGES (Project PB96-0797).

\end{document}